\documentclass[]{acmart}

\usepackage{graphicx}
\usepackage{tabularx}

\usepackage{listings}
\usepackage{verbatim} %
\usepackage{tikz} %

\usetikzlibrary[plotmarks] %

\newcolumntype{b}{X}
\newcolumntype{s}{>{\hsize=.3\hsize}X}

\usepackage{hyperref} 
\usepackage{todonotes}

\usepackage{graphicx}
\usepackage{subcaption}

\usetikzlibrary{positioning, shadows, shapes.geometric, shapes.symbols, patterns}
\newcommand*\pointer[1]{\tikz[anchor=2mm]{\node[shape=circle,fill=black, text=white,scale=0.5] (char) {\textbf{#1}};}}

\usepackage{xcolor}

\usepackage{csquotes}

\usepackage{listings}
\usepackage{upquote}

\usepackage{xparse}   %
\NewDocumentCommand{\rot}{O{45} O{1em} m}{\makebox[#2][l]{\rotatebox{#1}{#3}}}%

\definecolor{keyword}{HTML}{2771a3}
\definecolor{pattern}{HTML}{b53c2f}
\definecolor{string}{HTML}{be681c}
\definecolor{relation}{HTML}{7e4894}
\definecolor{variable}{HTML}{107762}
\definecolor{comment}{HTML}{8d9094}

\lstdefinelanguage{cypher}
{
	morekeywords={
		MATCH, OPTIONAL, WHERE, NOT, AND, OR, XOR, RETURN, DISTINCT, ORDER, BY, ASC, ASCENDING, DESC, DESCENDING, UNWIND, AS, UNION, WITH, ALL, CREATE, DELETE, DETACH, REMOVE, SET, MERGE, SET, SKIP, LIMIT, IN, CASE, WHEN, THEN, ELSE, END,
		INDEX, DROP, UNIQUE, CONSTRAINT, EXPLAIN, PROFILE, START,
	}
}

\colorlet{punct}{red!60!black}
\definecolor{delim}{RGB}{20,105,176}
\colorlet{numb}{magenta!60!black}
\colorlet{more}{green}

\lstdefinelanguage{json}{
	basicstyle=\normalfont\ttfamily,
	showstringspaces=false,
	breaklines=true,
	numbers=left,
	numberstyle=\scriptsize,
	stepnumber=5,
	numbersep=8pt,
	literate=
	*{0}{{{\color{numb}0}}}{1}
	{1}{{{\color{numb}1}}}{1}
	{2}{{{\color{numb}2}}}{1}
	{3}{{{\color{numb}3}}}{1}
	{4}{{{\color{numb}4}}}{1}
	{5}{{{\color{numb}5}}}{1}
	{6}{{{\color{numb}6}}}{1}
	{7}{{{\color{numb}7}}}{1}
	{8}{{{\color{numb}8}}}{1}
	{9}{{{\color{numb}9}}}{1}
	{<}{{{\color{more}<}}}{1} %
	{>}{{{\color{more}>}}}{1} %
	{:}{{{\color{punct}{:}}}}{1}
	{,}{{{\color{punct}{,}}}}{1}
	{\{}{{{\color{delim}{\{}}}}{1}
	{\}}{{{\color{delim}{\}}}}}{1}
	{[}{{{\color{delim}{[}}}}{1}
	{]}{{{\color{delim}{]}}}}{1},
}

\lstdefinestyle{ebnf}{language=sh,
	morekeywords={super},
	literate=
	*{0}{{{\color{numb}0}}}{1}
	{1}{{{\color{numb}1}}}{1}
	{2}{{{\color{numb}2}}}{1}
	{3}{{{\color{numb}3}}}{1}
	{4}{{{\color{numb}4}}}{1}
	{5}{{{\color{numb}5}}}{1}
	{6}{{{\color{numb}6}}}{1}
	{7}{{{\color{numb}7}}}{1}
	{8}{{{\color{numb}8}}}{1}
	{9}{{{\color{numb}9}}}{1}
	{:}{{{\color{punct}{:}}}}{1}
	{=}{{{\color{punct}{:=}}}}{1} %
	{;}{{{\color{punct}{;}}}}{1}
	{,}{{{\color{punct}{,}}}}{1}
	{\{}{{{\color{delim}{\{}}}}{1}
	{\}}{{{\color{delim}{\}}}}}{1}
	{[}{{{\color{delim}{[}}}}{1}
	{]}{{{\color{delim}{]}}}}{1},
}

\newcommand{\mycdots}{\cdot\!\cdot\!\cdot}
\lstdefinelanguage{javascript}{
  keywords={break, case, catch, continue, debugger, default, delete, do, else, finally, for, function, if, in, instanceof, new, return, switch, this, throw, try, typeof, var, void, while, with},
  morecomment=[l]{//},
  morecomment=[s]{/*}{*/},
  morestring=[b]',
  morestring=[b]",
  sensitive=true
}

\setcopyright{acmcopyright}
\copyrightyear{2023}
\acmYear{2023}
\acmDOI{XXXXXXX.XXXXXXX}

\acmConference[]{}{}{}
\acmPrice{}
\acmISBN{}

\begin{document}
\title{Towards Cross-Provider Analysis of Transparency Information\\ for Data Protection}

\author{Elias Grünewald, Johannes M. Halkenhäußer, Nicola Leschke, Frank Pallas}
\affiliation{%
  \institution{\\Information Systems Engineering, Technische Universität Berlin}
  \country{Germany}}
\email{\{eg, jh, nl, fp\}@ise.tu-berlin.de}

\begin{abstract}
Transparency and accountability are indispensable principles for modern data protection, from both, legal and technical viewpoints. Regulations such as the GDPR, therefore, require specific transparency information to be provided including, e.g., purpose specifications, storage periods, or legal bases for personal data processing. However, it has repeatedly been shown that all too often, this information is practically hidden in legalese privacy policies, hindering data subjects from exercising their rights. This paper presents a novel approach to enable large-scale transparency information analysis across service providers, leveraging machine-readable formats and graph data science methods. More specifically, we propose a general approach for building a transparency analysis platform (TAP) that is used to identify data transfers empirically, provide evidence-based analyses of sharing clusters of more than 70 real-world data controllers, or even to simulate network dynamics using synthetic transparency information for large-scale data-sharing scenarios. We provide the general approach for advanced transparency information analysis, an open source architecture and implementation in the form of a queryable analysis platform, and versatile analysis examples. These contributions pave the way for more transparent data processing for data subjects, and evidence-based enforcement processes for data protection authorities. Future work can build upon our contributions to gain more insights into so-far hidden data-sharing practices.
\end{abstract}

\maketitle 

\section{Introduction}

Transparency about the details of personal data processing is broadly considered an indispensable precondition for well-informed and self-sovereign privacy-related decisions and behavior. Only with relevant details on, for instance, the categories of personal data being collected and processed or the third parties they are to be transferred to being known, data subjects are able to consciously decide which services to use -- and how %
-- in the light of their personal, privacy-related preferences. Besides individual data subjects, similar information is also required by data protection authorities to fulfill their duties of, for instance, assessing the privacy risks emanating from single controllers, the types of personal data they collect, and the network of third parties they share these data with. For these tasks transparency information needs to be comprehensive and readily accessible. Fostering accountability, regulatory bodies need to be empowered to make informed decisions regarding the handling of personal data in complex scenarios \cite{karjalainen2022all}. 

Regulatory frameworks such as the European General Data Protection Regulation (GDPR) or the California Consumer Privacy Act (CCPA) therefore introduce obligations for data controllers to implement adequate transparency and accountability measures. These obligations particularly require data controllers to inform data subjects and data protection authorities about %
relevant aspects of the processing of personal data such as, for instance, the categories of data, the purpose(s) pursued, retention periods, third parties that personal data are transferred to, the employed legal bases, etc.%

So far, data controllers aim to achieve compliance with respective 
obligations through written %
privacy policies. This textual mode of providing transparency %
information implies a broad variety of well-known and well-researched shortcomings, ranging from general incomprehensibility for laypersons \cite{mcdonald2008cost, reidenberg2015disagreeable, trepte2015people} %
to paradigmatic incongruence with modern, highly agile development practices \cite{jensen2013towards, devprivops}. Besides these, however, it also severely hinders any attempt to perform higher-level analyses to be conducted on publicly reported transparency and accountability data, especially across multiple, and possibly interrelated, controllers.

If, for instance, data subjects could easily explore and assess the data sharing network of a service provider that unfolds behind legalese statements on third party transfers without having to laboriously examine the privacy policy of each and every mentioned third party (and, in turn, \emph{their} third party transfers), this would unquestionably heighten their informedness about what to expect from the use of the respective service. Individual data subjects may express satisfaction or contentment with certain levels of information disclosure regarding data processing practices. However, it is crucial to acknowledge that the requirements and needs of regulatory authorities extend beyond the expectations of data subjects.

Data protection authorities as well as policymakers, in turn, could also benefit from the revelation of such networks as well as from other higher-level analyses regarding, for instance, the use of different legitimating bases across different sectors, general and sector-specific distributions in the usage of different processing purposes or even simulations of the impact to be expected from two controllers merging into one during a competition-related approval process. Hence, regulators necessitate a more comprehensive and extensive scope of transparency information to effectively enforce regulations and ensure compliance with legal requirements. Thus, while data subjects may find certain transparency aspects pleasing, the regulatory context demands a much more exhaustive and thorough approach to transparency information. This also helps with the manifold tasks of data protection authorities, which include monitoring and enforcing compliance, providing guidance and advice to data controllers and data subjects, conducting investigations, and cooperating with others to ensure consistent application of the law across the scope of application (Art.~57~GDPR). 

Despite the promising value to be expected from such higher-level and especially cross-controller analyses being possible, we currently lack the technical capabilities and instruments for doing so. In this paper, we, therefore, propose an approach for facilitating such higher-level analyses based on legally obligatory transparency information being codified into a machine-readable format and integrated into a feature-rich transparency analysis platform (TAP). This TAP maintains an integrated, graph-structured representation of multiple providers' transparency information and allows applying state-of-the-art data science methods to, e.g., unravel data-sharing networks, analyze different processing scenarios, or try to predict the privacy-related effects of mergers %
across data controllers. More specifically, we:

\begin{itemize}
    \item propose a general approach for advanced transparency information analysis,
    \item design the architecture of a transparency analysis platform (TAP) leveraging machine-readable transparency information based on this approach,
    \item implement an open-source TAP featuring a rich set of functionalities and thereby supporting a broad variety of scenarios for automated data disclosure analysis, comprehension of data sharing networks, and evidence-based comparison of privacy policies.
    \item demonstrate practical applicability and provide illustrative examples with transparency information extracted from more than 70 real privacy policies, and
    \item simulate large-scale sharing networks and the effects of changes therein based on hyperparameterized synthetic transparency information.
\end{itemize}

Along these lines, this paper is structured as follows: In Section~\ref{sec:background} we provide relevant background and related work. We describe our general approach in Section~\ref{sec:general-approach}. Afterward, we describe how two analyze machine-readable transparency information in Section~\ref{sec:unravel}. We discuss our findings and conclude in Section~\ref{sec:discussion}.

\section{Background and Related Work}\label{sec:background}

In this section, we provide relevant background and related work. %

\subsection{Legal Transparency Obligations and Current Practices}
The principle of transparency in (privacy) regulations -- such as the General Data Protection Regulation (GDPR), the California Consumer Privacy Act (CCPA), or the %
Digital Services Act (DSA) coming into force soon -- deals with providing information to the data subject. Under the GDPR, for instance, respective obligations comprise, %
among others, the categories of personal data undergoing processing, well-specified purposes of the processing, applicable legal bases, aspired %
retention periods, information about third-country transfers, and many more (Art.~12--15~GDPR). All of these are meant to allow for well-informed decisions %
by data subjects on whether to use a specific online service or not. %

The established practice to align with the transparency principle is to provide a written privacy policy. However, repeated studies have shown that these are frequently not read due to their length \cite{jensen2005privacy, mcdonald2008cost, fabian2017large}, are hard to understand because of legalese \cite{reidenberg2015disagreeable} and/or technical \cite{tang2021defining} language, and often do not contain all necessary information \cite{yu2016can, bhatia2018semantic,torre2020ai}. In addition, they tend to include generalized and vague terms and therefore lack clarity \cite{reidenberg2016ambiguity}. Written privacy policies particularly fail in informing less privacy-literate data subjects about the data processing \cite{trepte2015people}, but also experts have difficulties interpreting these policies \cite{reidenberg2015disagreeable}.
Overall, this leads to data subjects being unable to evaluate the risks of using a service \cite{chang2018role}. Current practices thus broadly fail to address the actual purpose of regulatory obligations.

\subsection{Transparency Enhancing Technologies} %
To overcome these and further shortcomings of merely textual transparency information, different approaches for technically mediated transparency have been proposed. Large parts of respective research can be subsumed under the term of \emph{transparency enhancing technologies (TETs)}. Such proposed TETs range from low-level server-side technologies for collecting transparency-relevant details of the processing \cite{angulo2015} over microservice annotations explicitly tailored to regulatory requirements \cite{grunewald2021tira} to user-facing approaches such as in-browser cookie tracking \cite{lightbeam} or privacy dashboards \cite{bier2016privacyinsight, Raschke2018, tolsdorf2021case}. More recent forensic documentation systems, such as the powerful webXray, are more aligned to specific legal needs but inspect privacy violations also from a pure client perspective \cite{libert}. In matters of communicating privacy- and especially transparency-related information in a manner more comprehensible than the above-mentioned written privacy policies, proposed approaches include easy-to-remember visual representations \cite{fischer2016transparency}, such as privacy icons \cite{holtz2011towards, iannella2010privacy, rossi2020, togglesDollar2021} or serious comics \cite{anaraky2019testing}. Layered approaches integrating iconic symbols with summarizing text statements \cite{grafenstein2021effective, gluck2016short} have also been discussed recently. TETCat~\cite{tet_cat_zimmermann} provides an exhaustive categorization of the different kinds of TETs.

A specific, yet often overlooked subfield of research on technically mediated transparency regards the machine-readable representation of transparency information in specifically tailored formats and languages, which is indispensable for rendering the above-mentioned approaches practically viable and scalable. In addition, having the underlying transparency information represented and available in such machine-readable form also allows to automatically generate different presentations or even inherently versatile user interfaces in a %
context-, preference-, or competence-adaptive manner \cite{gruenewald2021}. %
Early examples of such languages particularly include the Platform for Privacy Preferences Project (P3P), which encoded some static data categories and purposes \cite{cranor2003p3p}.
To meet the requirements of modern privacy regulations, however, a much more comprehensive expressiveness is required. Recent proposals that consciously incorporate these obligations include the Layered Privacy Language (LPL) \cite{gerl2018lpl}, or the Transparency Information Language and Toolkit (TILT) \cite{gruenewald2021}. With these, data controllers are able to codify their %
data processing activities and to exchange the stated information through various channels. Related work provides a general overview of %
respective formats available \cite{leicht2019survey}. %

Exceptions like (so far not technically implemented) design concepts for illustrating sharing networks \cite{grafenstein2021effective, efroni2019privacy} notwithstanding, most of these approaches do, however, only focus on an isolated, per-controller perspective and do not pay proper regard to cross-provider issues.

\subsection{Cross-Provider Analysis of Privacy-Related Practices} As pointed out above, cross-provider issues are of significant importance for \enquote{getting the whole picture} of a provider's privacy-related practices and the associated risks and implications -- from the perspective of data subjects as well as authorities and regulators.

Related work in this regard primarily focuses on understanding to what extent controllers comply with regulations, on how the communication and accessibility of privacy policies have changed over time, and on analyzing the content of privacy policies more broadly \cite{krisam2021dark, amosPrivacyPoliciesTime2021, gotze2022}. To understand how controllers comply with regulatory givens, scholars look at how well correct legal bases have been provided \cite{wagnerPrivacyPoliciesAges2022} or to what extent data-sharing has been appropriately documented. Employed methods and techniques here include natural language processing (e.g. \cite{harkousPolisis2018, zimmeckMAPS2019}), human annotation \cite{wilsonCreation2016}, or automatic evaluation of cookies \cite{bollinger2022automatingCookieConsent, thotawaththa}.

With a particular focus on the sharing of personal data across data controllers and the structure of such networks, related work is rather sparse. For example, Urban~et.~al. analyze the development of data-sharing and the relative changes in the resulting network structure before and after the enactment of the GDPR \cite{urban2020adNetworks}. However, they conduct their analysis solely based on monitored %
client-side cookies. Similarly, \cite{amosPrivacyPoliciesTime2021} also only discusses third-party sharing with respect to cookies shared across controllers. %
Although the presented insights are of great value (e.g., networks became more centralized as the GDPR took effect), this approach does not include data-sharing that happens only in the control sphere of the controller or, in other words, in the \enquote{backend} infrastructure. In addition, cookies do not contain other necessary information, such as the legal basis for the data disclosure. 
Insofar, novel approaches are needed that also cover activities of controller-side (backend) data sharing and additionally pay proper regard to further aspects deemed relevant from the regulatory perspective, such as the purpose of or the legal basis employed for sharing personal data with third parties. As data is stored and processed increasingly in distributed computing environments, data controllers must navigate the complex landscape of transparency and accountability requirements to ensure the protection of personal data and other sensitive information. In response to these challenges, innovative approaches to transparency and accountability are needed.

In matters of analytical methods, 
graph analysis is broadly established for network-focused topics \cite{wohlgemuth2014facebookNetworks, knoke2019social} and is increasingly used for addressing privacy-related aspects in the light of significant cross-provider interdependencies. %
For instance, related work uses machine learning models to harness the graph properties of websites for blocking tracking cookies \cite{iqbal2020adGraph}. Common analysis techniques here include the clustering of nodes, estimating and quantifying the relationships between different nodes, calculating the centrality of a network, link prediction, dynamic simulation of graph developments, and many more. With the advent of modern graph database features, such as those built into Neo4j, these techniques become more and more accessible \cite{needham2019graph}. Therefore, these are well-suited especially for the analysis of inherently complex data-sharing activities, as we will explain below.

\section{General Approach}%
\label{sec:general-approach}

Based on the current state of related work and the hitherto unsatisfied needs identified above, we propose a novel \textit{transparency analysis platform} to unlock the so-far isolated and controller-specific representations for in-depth analysis across multiple controllers, which could potentially be numerous. Our general approach to achieving this goal involves a workflow that begins with the machine-readable codification %
of transparency information for a multitude of controllers with potential data-sharing interrelations. %
Subsequently, we investigate and apply different possibilities for conducting meaningful graph analysis on said transparency information, using a prototypical implementation of our envisioned transparency analysis platform. %
We thereby show how graph analysis of transparency information enables us to better understand the relationships between different controllers and to identify patterns and trends that would be otherwise difficult to discern manually from written privacy statements.
In light of the feature-rich language design, the accompanying programming libraries, and the preexisting document corpus, we herein opt for the above-mentioned TILT as the core technology for our %
considerations, albeit without loss of generality for other %
 formats \cite{leicht2019survey}.\footnote{We invite data controllers, organizations, and individuals to participate in a community effort by providing machine-readable transparency information in any interoperable format.} %

Proposing this workflow, we aim to provide deeper insights into the %
actual data-sharing practices of controllers that are hardly recognizable by data subjects and other third parties so far. On this basis, we aim to achieve a better level of sharing-related transparency for data subjects and to support the development of more effective and efficient enforcement strategies for data protection authorities.%
Last but not least, we also aim to uncover and shed light on %
complex data-sharing networks in general, merely employing legally obligatory and, thus, publicly available transparency statements. In more detail, our %
approach unfolds as follows:

\textbf{1. We collect transparency information from real-world controllers to be provided in a machine-readable format.} For doing so, we publish the collected information in a public open-source repository of TILT documents.\footnote{\url{https://github.com/Transparency-Information-Language/tilt-corpus}} %
The corpus of machine-readable transparency information were extracted from real-world online. They contain information about the collection, processing and use of personal data in accordance with the legal requirements of the GDPR. This collection covers more than 70 controllers from different industries, such as social media, e-commerce, or online banking, news websites or higher education institutions across Europe. For this work, we particularly enhanced the documents to be complete regarding the transparency obligations from Art.~12--15~GDPR. Additionally, for each covered controller, we added an industry sector classification following the International Standard Industrial Classification of All Economic Activities (ISIC) proposed by the United Nations for later comparisons across business domains. 

\textbf{2. Next, we define a graph structure comprising all information elements the schema consists of.} The respective schema results from the structured respresentation in place. If using TILT, several nodes are to be created per controller %
with each %
being dedicated to one relevant category of personal data and holding the respective transparency information such as 
purpose, legal bases, storage periods, subject rights, and receivers, possibly resulting in third-country transfers. We then transform the information of all present TILT documents into this graph structure. Moreover, we connect %
said nodes through edges that represent explicit data-sharing statements from
the original privacy policies.
A similar construction would be possible using the elements defined as tuples (e.g., purpose, entity, data source, data recipient etc.) in the Layered Privacy Language \cite{gerl2018lpl}.

Moreover, we compute similarities between nodes based on certain attributes such as, e.g., the name of the data controller to allow for automated node-matching. These similarities are based on common similarity metrics, in particular, %
the Jaro–Winkler distance, the Sørensen–Dice coefficient, and the Levenshtein distance, which are all common to quantify string similarities \cite[e.g.,][]{vijaymeena2016survey}. All the resulting nodes and edges of the graph structure are stored in the dedicated graph database (Neo4j). Remarkably, the graph database offers various graph data science algorithms for direct application to the stored data (e.g., the popular Page-Rank algorithm, or several clustering techniques). 

\textbf{3. Our general architecture, moreover, comprises a queryable API.} Using GraphQL, queries can be executed to create, read, update, or delete (CRUD) nodes and edges in the graph database. Through the usage of the GraphQL query language, we prevent the challenges of over- and under-fetching. Instead, the API consumers can query for the exact transparency information elements they need for their analysis task. Consequently, queries can now not only provide information about a single controller but also about connected data-sharing networks.
In addition, we support the use of the native graph database query language \textit{Cypher} in Neo4j. 

\textbf{4. Furthermore, we can access the data through the provided API or the query language. Ultimately, we can visualize the results of all queries using various plotting libraries or the recommended graph database dashboards.} To this end, we can leverage existing algorithms to optimize the presentation of structured graph representations. Through the provided API, these visualizations can be viewed in a multitude of contexts, depending on the preferences and existing tools of the legal entity at hand. Some transparency information representations might allow for a more detailed view (e.g., TILT), while others rather focus on much more aggregated information (e.g., TOS;DR\footnote{\url{https://tosdr.org}}).
The resulting visualizations can be integrated into existing privacy dashboards or be used to create new ones. Others means of communication can also be utilized \cite{gruenewaldEnablingVersatile}, such as privacy icons. Some supervisory authorities might even more comprehensive analysis needs, especially to safeguard information in very large data ecosystems, such as governmental authorities and their subordinate institutions. In this case, the data could also be exported and analyzed in a dedicated environment.

\textbf{5. Next, we provide means for generating synthetic transparency information to validate and illustrate the viability of our approach and the value that could be expected from a broad availability of machine-readable transparency information.} Our generator offers two different approaches for doing so, with the first one being based on the observed statistical distributions of the existing documents (e.g., most frequently utilized legal bases, the average period of storage, or sharing relationships to controllers operating in a similar industry sector). Alternatively, the synthetic information can also be generated using hyperparameters that are arbitrarily selected for these elements. These data can be instrumental in validating the efficacy of our approach and demonstrating the value of machine-readable transparency information for comprehensive analysis and decision-making processes in general.
    
\textbf{6. Given a potentially large corpus of machine-readable transparency information in a queryable graph structure, we can simulate possible network dynamics using suitable hyperparameters.} These include, for instance, the likeliness of a merge of two controller nodes (read \enquote{company $A$ buying company $B$}) in the modelled graph structure. This also enables analysts, e.g., in supervisory authorities, to evaluate the associated risks of the resulting (merged) data-sharing activities, which is a crucial aspect of the accountability principle. Other examples, relevant for supervisory authorities, include identifying potential correlations of personal data among a few entities, the emergence of new data-sharing networks, the detection of trends of inaccurate purpose specifications or alike.       

In short, our general approach allows for the stringent analysis of transparency information in a suitable machine-readable format, such as TILT. This is achieved by transforming the information into a graph structure and providing a queryable API. The graph structure allows for the application of graph data science algorithms, which can be used to analyze the data-sharing networks of controllers.

\subsection{Architecture and Implementation}

\begin{figure}[t]
    \includegraphics[width=0.9\linewidth]{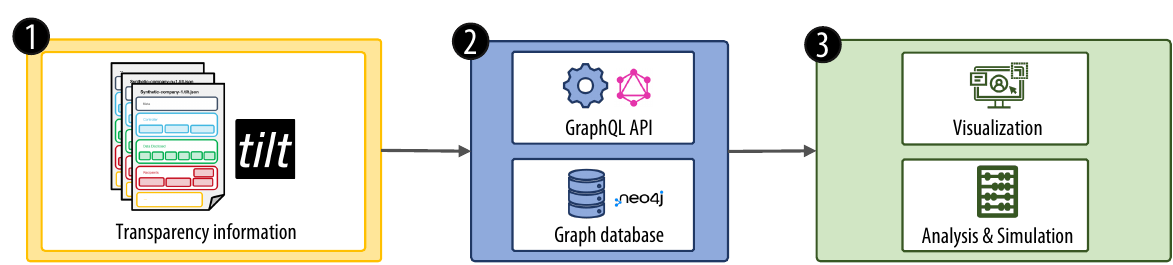}
    \caption{General architecture of the Transparency Analysis Platform.}
    \label{fig:implementation}
\end{figure}

Our general architecture is depicted in Figure~\ref{fig:implementation}. Considering the general approach introduced above, we \pointer{1} assume relevant transparency information to be present in machine-readable form, such as TILT. This information could have been extracted from privacy statements or has been generated from other transparency enhancing technolgies. At the core of the transparency analysis platform, this information is then transferred to the \pointer{2} graph database, for which we choose Neo4j. Access to the data can be made through the powerful yet flexible GraphQL API, or the established database query language Cypher. Lastly, the information can be \pointer{3} visualized, statistically analyzed, or used to perform simulations. Furthermore, through containerization technologogies, the application allows for easy deployment in organizations and version control with low integration efforts. Altogether, this allows users, especially from supervisory authorities, to flexibly interact with and further integrate the data into various contexts.

\subsection{Transparency Information Graph Schema}

\begin{figure}[t]
\centering
    \includegraphics[width=0.9\linewidth]{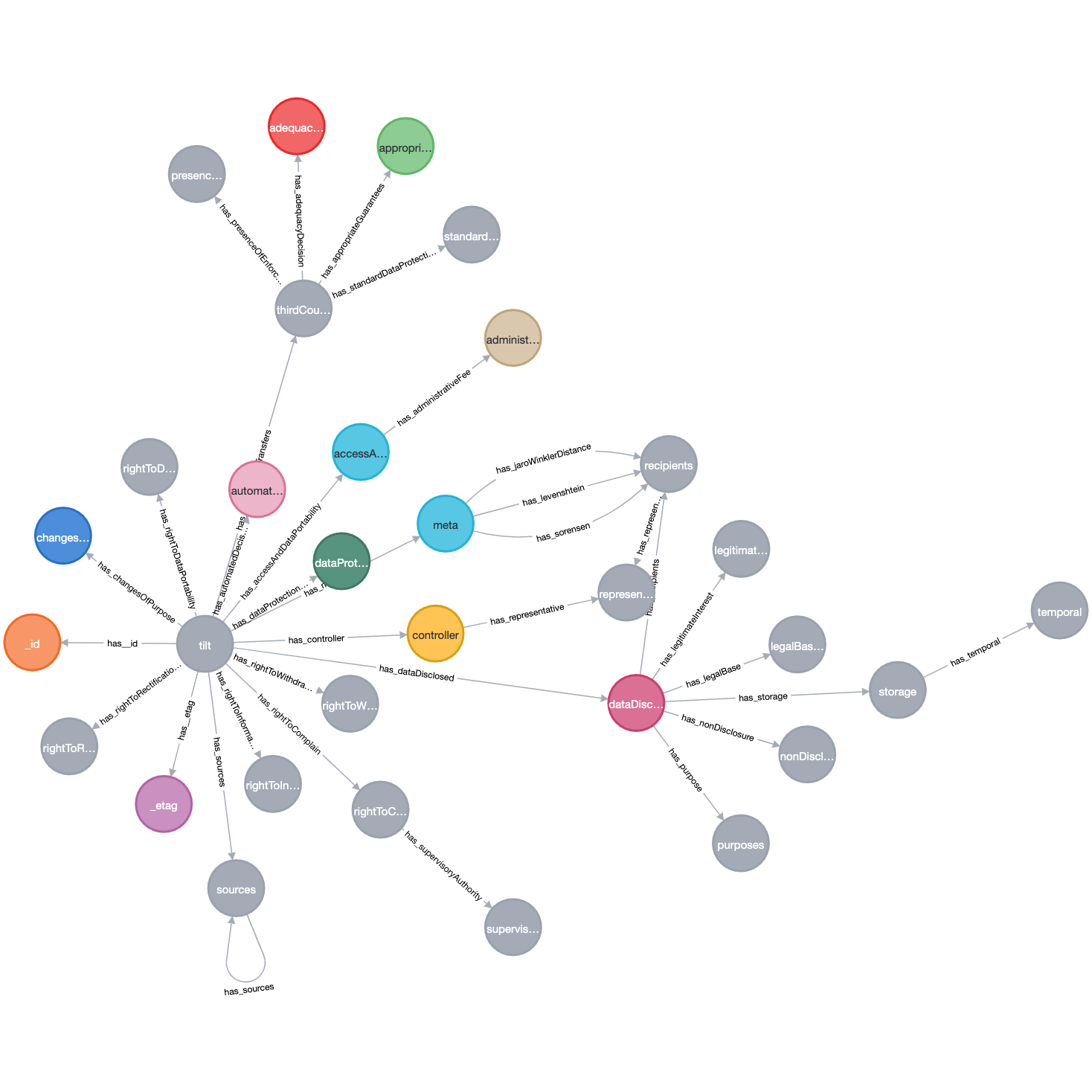} 
    \caption{Transparency Information Graph Schema.} %
    \label{fig:graph-schema}
\end{figure}

The original TILT representation\footnote{Find the full specification here: \url{https://transparency-information-language.github.io/schema/}}, encoded in JSON Schema, allows for a graph structure representation where links between documents are made explicit and exploitable for further analysis. Hence, we converted the TILT documents into the graph schema depicted in Figure~\ref{fig:graph-schema}. Allowing for the greatest possible flexibility in the analysis, we develop a versatile recursive function to read and upload TILT documents into Neo4j using the GraphQL API based on an explicitly defined GraphQL schema. The function builds a GraphQL query using %
each element of the TILT format as its base-case. The more than 70 pre-existing %
TILT documents are hence transformed into separate sub-graphs, with each part of the schema being a new node. Any non-dictionary entry becomes a queryable attribute of a node. With the embedded Neo4j graph data science library, we perform community or link detection, which we will elaborate on below. 

Within the schema, the meta nodes contain core information about the data controller. Hence, any inferred relationship between individual graphs will be rooted to the meta nodes. %
On this basis, the user %
can construct personal data-sharing networks with a more global view than isolated analysis of a single policy, which allow for further analysis and understanding of which companies are actually interconnected.
As we will also see below, nodes can be related through data-sharing, similar-looking purposes for disclosing data, or categories of personal data being processed, by the underlying logic.
In a nutshell, the relationships can be constructed by the users using a fuzzy match of controllers' names or text similarity (with the similarity measures mentioned in section~\ref{sec:general-approach}). The threshold above which two entries are considered similar enough to qualify for a match is hyper-parameterized, allowing for versatile analyses.

\section{Analyzing Machine-Readable Transparency Information}
\label{sec:unravel}

Based on the introduced general approach, presented architecture, and implementation, we can harness a rich set of functionalities supporting the disclosure analysis and (user-side) comprehension of data-sharing networks. Respective illustrative analysis examples shall be elaborated on in the following. 

\subsection{Queryable Views}\label{sec:views}

The transparency analysis platform allows for a multitude of different views on the data at hand since all properties of nodes and relationships between them can be queried. Correspondingly, the user can get insights into all desired pieces of transparency information of an isolated single controller, a set of controllers that are linked through explicitly stated data-sharing practices (see below), or the global perspective of all data controllers contained in the pool. %
Leveraging the standardized machine-readable format, we can now analyze particular aspects of interest: Giving a first illustrative example, we want to understand whether different data controllers use different legal bases to justify disclosing data to third parties depending on their industry sector. For doing so, we run the following Cypher query\footnote{In this paper, we omit to explain the general syntax of Cypher queries, but point the interested reader to the official documentation: \url{https://neo4j.com/developer/cypher/}} against the graph database interface:

\begin{figure}
    \begin{lstlisting}[language=cypher, label=lst:legalBases, caption=Comparing legal bases across industry sectors.]
    MATCH (m:meta)-[]-(t:tilt)-[]->(d:dataDisclosed)-[]->(l: legalBases), (t)-[]->(c:controller)
    RETURN m.name, c.sector, d.category, l.reference
    \end{lstlisting}
\end{figure}

\begin{figure}[t]
    \centering
    \includegraphics[width=0.55\linewidth]{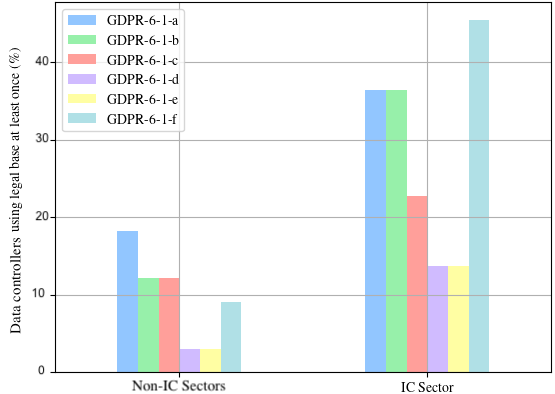}
    \caption{Comparing the legal bases for data controllers along the Information and Communication (IC) sector, citing a legal base at least once. For instance, GDPR-6-1-a refers to \textit{consent}.}
    \label{fig:bases}
\end{figure}

The query yields all nodes that hold the respective information, which are any categories of data disclosed and the specified legal basis. Since the GDPR most prominently differentiates between six possible legal bases, as expressed by Art.~6(1)~lit.~a,~b,~c,~d,~e, and~f, 
we can now provide valuable insights on the actual distribution of these in real-world scenarios. Figure~\ref{fig:bases} shows the results for the query, with a further distinction %
between the Information and Communication (IC) and other sectors. From our dataset, we learn that controllers classified as IC are generally using more legal bases as justification for their processing activities, while they also massively make use of the Art.~6(1)~lit.~f alternative.\footnote{Nota bene: Such an illustrative example is, like the following ones, not claiming for generality, since the dataset is limited to 70 data controllers. Rather, we want to demonstrate the novel general way to perform such queries of interest.} This refers to the processing based on legitimate interests, which is controversially discussed in literature \cite{kamara2018understanding}, because it is often used as a catch-all legal basis. 

Henceforth, all subsequent analytical queries %
follow the same approach. First, a query is made against the graph database, and second, the results may be summarized and visualized using general-purpose libraries. In particular, we use standard Python libraries and publish all our explained queries (and some more) in a public repository.\footnote{\url{https://github.com/Transparency-Information-Language/tilt-graph-analysis}}   

\subsection{Entity-Matching and Linkage Creation}

When extracting transparency information from real-world privacy policies, we identify a common problem: Data controllers mentioned as receivers of data disclosed are inconsistently denoted. For instance, \enquote{Firm $A$}, \enquote{$A$ Ltd.}, or simply \enquote{$A$} should all relate to the same legal entity. Hence, a na\"ive query cannot match these entities exactly to one another. Instead, we utilize a similarity algorithm to calculate the distance between two identifiers. If this distance is %
below a parameterized threshold, the two entities are considered to be the same legal entity. The most common similarity algorithm is the Levenshtein~distance, which is a recursive function that considers the length of the word and how many times a letter needs to be replaced, inserted, or removed to transform one word into another. However, the Sørensen-Dice similarity is preferred for fuzzy matching compared to the Levenshtein~distance, which works well for finding minor spelling differences. The Sørensen-Dice coefficient calculates the ratio between the common elements of two words and the length of the words:

\[ \textrm{DSC} = \frac{2|X \cap Y|}{|X| + |Y|} \]

After such pre-processing, data controllers but also legal bases or purposes can be matched. Overall, using several such similarity measures allows us to connect nodes of a controller to, e.g., a data-disclosure node from another data controller and thus extract a network based on a variety of connections between those. Ultimately, the user can control which of these connections shall be used for further analysis. This way, we allow for comparing all types of transparency information, even in case of orthographic differences. A simple query yielding nodes and recipients with a high Sørensen-Dice similarity can be queried and directly connected as depicted in listing~\ref{lst:sorensen}:

\begin{figure}[!h]
    \begin{lstlisting}[language=cypher, label=lst:sorensen, caption=Connecting data controllers and recipients based on identifier similarity.]
MATCH (m:meta), (r:recipient)
WHERE apoc.text.sorensenDiceSimilarity(apoc.text.clean(m.name),
      apoc.text.clean(r.name), 'en') > {threshold} 
MERGE (m)-[c:has_sorensen]->(r)
    \end{lstlisting}
\end{figure}

In general, such matching techniques allow for efficient search across the whole data set. Compared to a manual process, in which a person had to read and extract all relationships from numerous privacy policies, underlines the utility of our prototype.

\subsection{Clustering and Centrality Measures}

To unravel data-sharing networks previously hidden to data subjects or supervisory authorities, our prototype allows for the clustering of data controllers based on all available transparency information. This can yield further, highly valuable insights, as it can be exemplified with the following type of conducted analysis. %

The Louvain algorithm %
\cite{blondel2008unfoldingCommunities}, for instance, %
creates clusters while optimizing
the modularity (i.e., the strength of division) of the entire network: First, the modularity of the graph is optimized by changing an individual node's membership within a cluster. The change in modularity is calculated using

\[ \Delta Q(D \rightarrow i \rightarrow C) = \Delta Q(D \rightarrow i) + \Delta Q(i \rightarrow C) \textrm{,} \]

where $Q$ is the modularity, $D$ is the original community, $i$ is the node in question, and $C$ is the target community. Second, after all nodes have been assigned to a community based on the modularity optimization, the communities are grouped and form \enquote{super nodes}, which can be connected as well. %
The algorithm is repeated until a threshold of modularity or a number of communities has been identified. Before starting the algorithm, a prior can be passed to it, that initializes which communities the individual nodes belong to, which can in our case be helpful controllers signaling such connections themselves or supervisory authorities possessing additional prior knowledge.

\begin{figure}[t]
    \centering
    \includegraphics[width=0.50\linewidth]{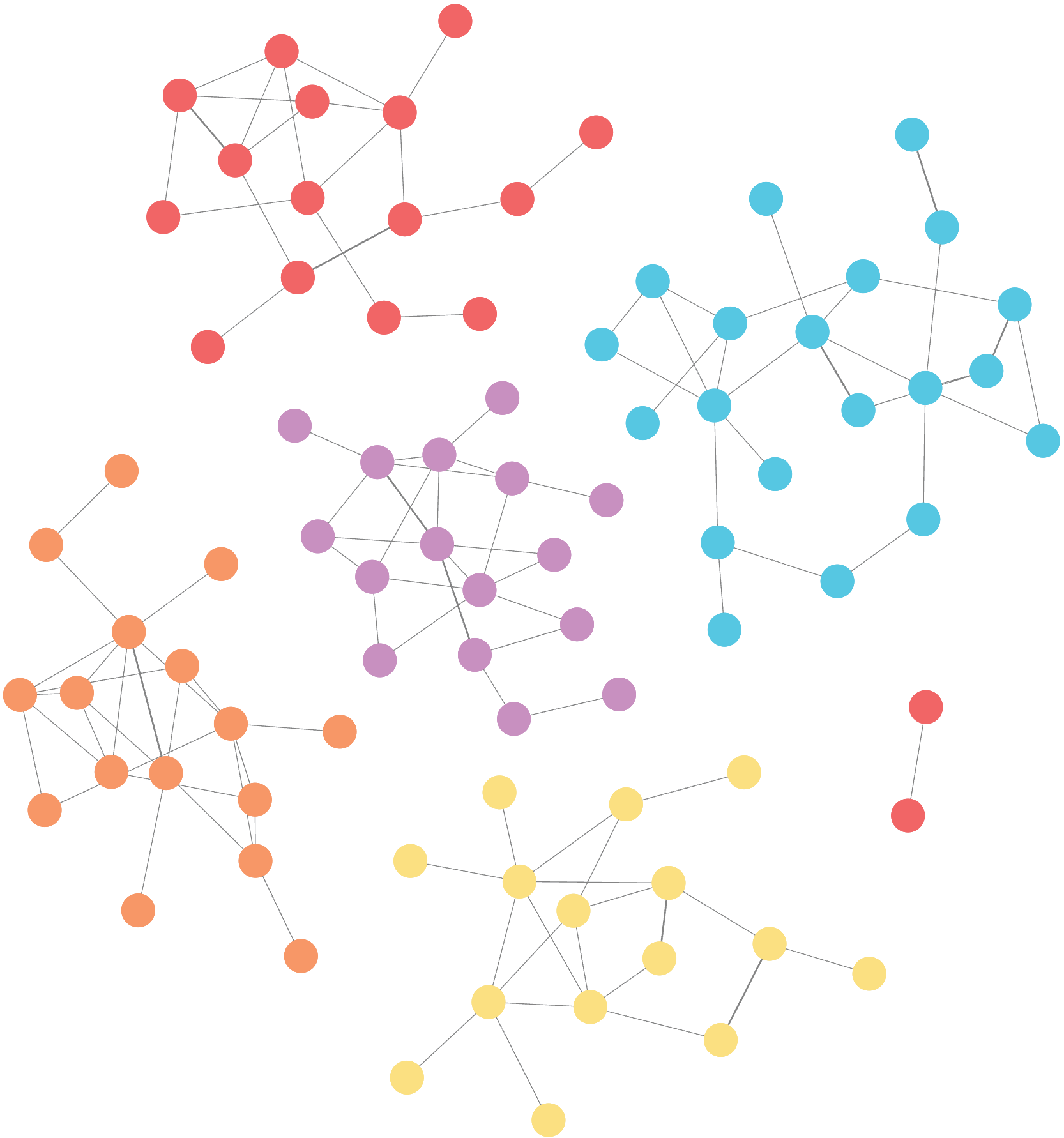}
    \caption{Exemplary resulting clusters based on Louvain community detection.}
    \label{fig:louvain}
\end{figure}

We argue that the inferred communities (or so-called \enquote{clusters}) can serve to find subnetworks that share personal data most among each other, yielding a more transparent understanding of the underlying network structure. They also allow for an assessment of whether treatment effects of (new) regulatory givens can be understood through community effects. For example, if a regulatory obligation is applied to individual controllers due to their size (e.g., the EU Digital Services Act only targets Very Large Online Platforms, VLOPS), is there an effect on smaller controllers in the same community and if so, can it be predicted by community status? In Figure~\ref{fig:louvain}, we show how the Louvain algorithm would cluster our exemplary dataset after linkage prediction as explained above. Likewise, the TAP supports other clustering algorithms enabled by the Neo4j Graph Data Science (GDS) library.

Moreover, we apply centrality measures to the data-sharing communities based on the transparency information at hand. Among others, we used the PageRank algorithm \cite{langville2011google} to associate nodes in the communities with different weights. The PageRank algorithm defines the influence of a node within a network by taking into account all its incoming and outgoing edges (relative importance). In interconnected scenarios, in which many data controllers transfer personal data to the same (often very few) data recipients, these can potentially aggregate these data to infer so far isolated knowledge. Consequently, the associated risks of transferring personal data to a third party with a high centrality value should be reasonably justified. With such analyses, we provide a tool to better evaluate which nodes should therefore be under increased attention. In our sample documents, the centrality analysis using PageRank showed a very skewed distribution of few (and well-known) %
controllers with outstanding centrality. With %
more comprehensive 
underlying datasets, the TAP can 
expectably unravel such highly influential data recipients and providers even more clearly and completely.

Based on the examples described, it can be inferred that simply providing transparency information -- without additional analysis -- has not been sufficient to achieve a practical level of understanding. 
We demonstrate that this is particularly relevant in such interconnected settings we observe. Thus, our approach can guide future research in this direction, which would be particularly helpful with a representative %
amount of machine-readable transparency information at hand, optimally provided by data controllers themselves. In general, such clustering techniques are a method to automatically identify data-transfer relationships apart from the ones explicitly codified in the underlying data. For instance, by understanding which nodes are central to a network, regulators and auditors can understand which data controllers are critical to the functioning of the environment, where to regulate, or where a security breach/data leak would be most harmful. 

\subsection{Network Dynamics} 

\begin{figure}[t]
  \centering
  \begin{subfigure}{0.32\linewidth}
    \includegraphics[width=\linewidth]{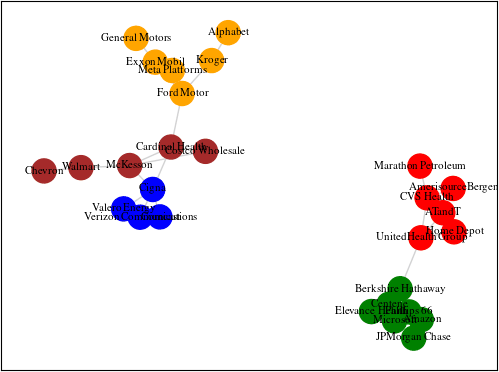}
    \caption{$t=2$}
  \end{subfigure}
  \begin{subfigure}{0.32\linewidth}
    \includegraphics[width=\linewidth]{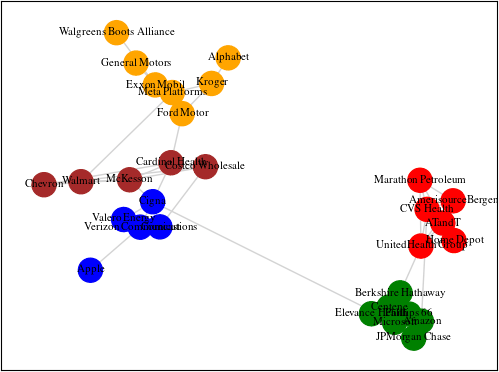}
    \caption{$t=41$}
  \end{subfigure}
  \begin{subfigure}{0.32\linewidth}
    \includegraphics[width=\linewidth]{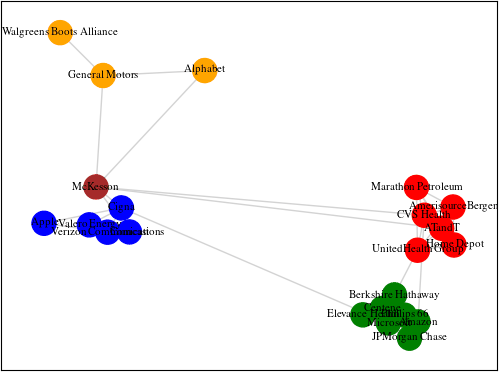}
    \caption{$t=94$}
  \end{subfigure}
  \caption{Example of data-sharing network dynamics with several merging events after $t$ iterations. Note, the depicted data controllers and their sharing activities are not based on real data but only illustrate the potential of network dynamic simulations.}
  \label{fig:dynamics}
\end{figure}

Cross-provider analyses of transparency-related practices also need to take into account network dynamics over time. Related work provided the first large-scale longitudinal analysis results, considering key term trends in privacy policies, word count, or readability scores \cite{amosPrivacyPoliciesTime2021}. However, these types of text analysis cannot automatically depict changes with the necessary granularity and semantic meaning. The automated extraction of processing activities, in turn, is a challenging and so far largely unsolved task \cite{bui2021automated}. In particular, respective studies often lack the necessary precision to reliably return actual legal bases, purpose descriptions et cetera. 

Moreover, these static analysis methods do not allow for simulating common real-world scenarios, such as the merging of two (or more) data controllers. In case such a merge is planned, supervisory authorities lack the tools to independently make evidence-based assumptions about probable sharing-network clusters or centrality changes. Frequent examples with relevance to privacy include merger and acquisition events, in which two or more legal entities also aim to merge their respective portfolios of (personal) data. 

With our TAP, in turn, we provide a tool to simulate and evaluate such network dynamics with suitable hyperparameters. In our experiments, we can scope a set of controllers and simulate certain events, based on different probability distributions. As shown in Figure~\ref{fig:dynamics}, in several iterations (allowing for longitudinal studies) edges are added to controllers based on, e.g., their centrality or a chosen category of personal data disclosed. After each iteration, we re-compute the clusters as described above. Following another probability distribution, we merge nodes representing acquisition events. Eventually, we can execute queries of interest as described in section~\ref{sec:views}. This allows the user of the TAP to compare, e.g., the number of data categories disclosed, the distribution of third countries involved, the frequency of specific purpose specifications, and many more.

In future work, these kinds of network analyses can also take into account historical data to perform predictions or trend analyses. We also emphasize the flexibility to choose the hyperparameters (see next section for some examples) to simulate data-sharing activities across service providers to, for instance, model scenarios of different likelihood. This can be of particular interest for supervisory authorities to evaluate the associated risks of data-sharing activities.

\subsection{Synthetic Transparency Information}

The real-world data at hand, although of great value to demonstrate the above-mentioned analysis techniques, do not provide for a densely populated data-sharing network. Only a few controller pairs could be matched to be sharing data with one another. Given such a sparsity of the adjacency matrix, statistical analysis cannot be above common significance levels. To simulate %
what would be possible with the broad availability of machine-readable
transparency information, we %
generated 
a separate, 
synthetic
dataset containing a variable number of controllers connected to their respective data disclosed and purpose nodes whose number is generated based on a hyperparameterized Poisson distribution $P(N_d) \sim Poisson(\mu)$,  where $N_d$ is equal to the number of data disclosed nodes to be generated for a controller. Each node is assigned to an underlying cluster (which can be based on the industry sector classification). The likelihood with which a data disclosed node (containing the information with whom data is shared) is connected to another controller is hyperparameterized and can be made dependent on whether two nodes share the same cluster. When a node is selected, the nodes in the same cluster are hence weighted higher.

    \begin{figure}[!t]
    \centering
    \includegraphics[width=0.75\linewidth]{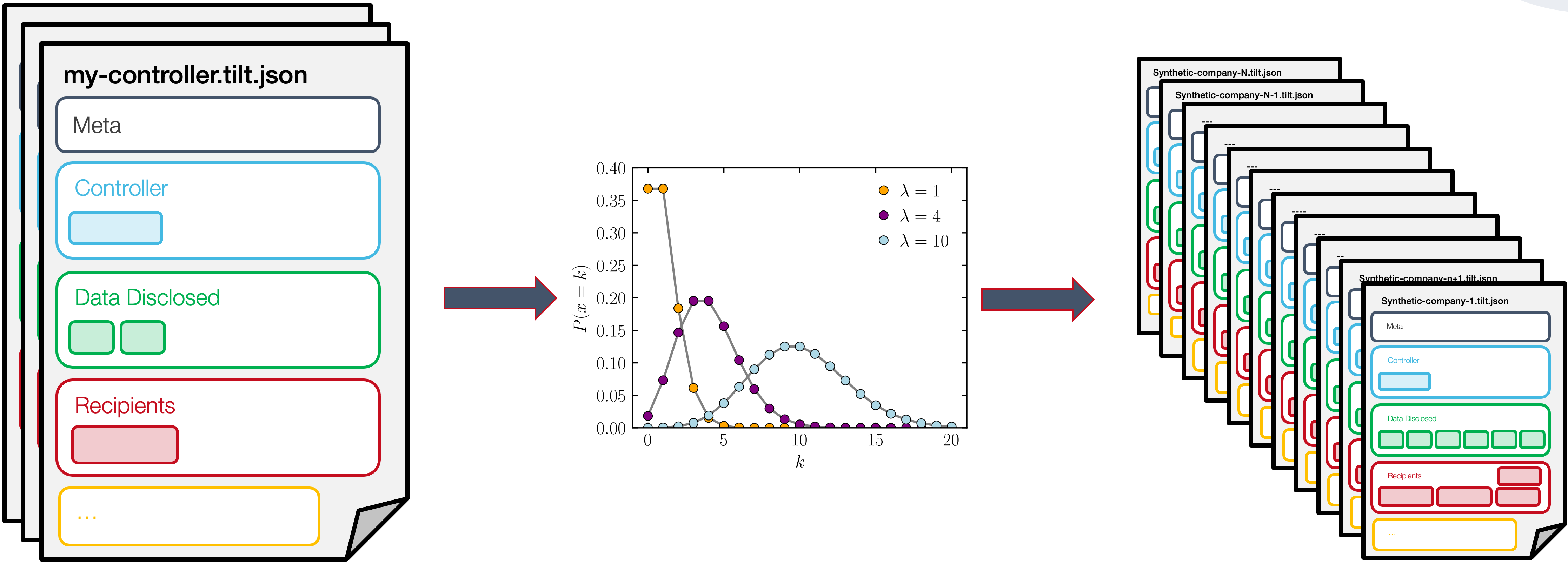}
    \caption{Based on the distribution information from the original TILT dataset, we can, e.g., induce how many data items are disclosed on average per controller or which legal bases are indicated and use these distributions for generating synthetic data. Fig.~based on: \cite{poisson}}
    \label{fig:synth}
\end{figure}

Subsequently, we can simulate various other cluster densities in the setup of the graph. In particular, we can generate synthetic transparency information on which to conduct further analysis based on the implicit distributions in our dataset at hand (cf.~Figure~\ref{fig:synth}).
Generating synthetic data based on real distributions leads to more realistic outcomes than relying on purely stochastic distributions.

\begin{figure}[t]
  \centering
  \begin{subfigure}{0.4\linewidth}
    \includegraphics[width=\linewidth]{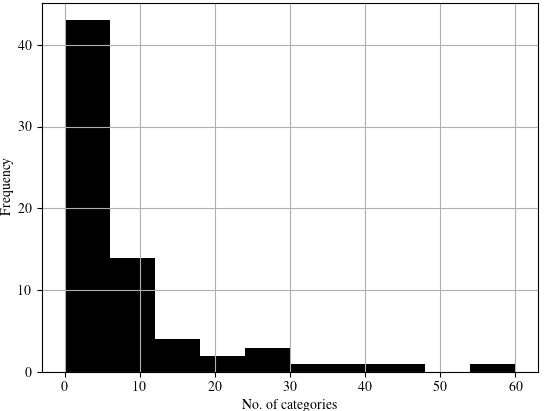}
    \caption{Data categories per controller.}
    \label{fig:dist-cat}
  \end{subfigure}
  \begin{subfigure}{0.4\linewidth}
    \includegraphics[width=\linewidth]{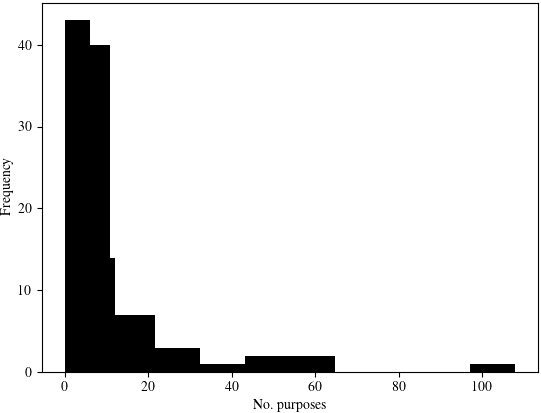}
    \caption{Purpose specifications per controller.}
    \label{fig:dist-purp}
  \end{subfigure}
  \caption{Observed distributions of selected transparency information in our dataset, which can be modeled with different statistical distributions.}
  \label{fig:distributions}
\end{figure}

In Figure~\ref{fig:distributions}, we show selected distributions of transparency information observed. Considering the number of data categories per data controller in Figure~\ref{fig:dist-cat}, we measure a mean of $\mu = 7$ categories per controller. Likewise, in Figure~\ref{fig:dist-purp} we can query for the mean of purpose specifications per controller, which results in a mean of $\mu = 8$ (both including the data controllers that have no specification at all). Such results can be of interest for future policy work, but can then also be used as hyperparameters for the Poisson distribution mentioned before to create more realistic datasets. Thereby, we are not limited by the amount of machine-readable transparency information available already regarding the active development and extension of our platform.   

Eventually, such synthetic datasets can be utilized to create more and more controlled experiments for applying graph data science methods to privacy-relevant information. We, therefore, pave the way for more advanced transparency-related analyses. This can help to fairly evaluate the data-sharing practices of different data controllers and clusters thereof, e.g., regarding the associated risks of disclosing personal data in the first place. Ultimately, we expect transparency information to be available from many more real-world data controllers due to regulatory obligation or trust-building signalling activities (i.e., providing the information voluntarily) from data controllers. Of special interest are also transparency information in other expressive formats, such as the IAB Europe Transparency and Consent Framework (TCF) \cite{iab}.

\section{Discussion, Future Work, and Conclusion}\label{sec:discussion}

With our contributions, we show the previously underappreciated potentials of coherent graph-based analysis of transparency information. Using our extensible TAP, we simplify query processing, yielding results for both specific controllers or even across implicit or explicit data-sharing networks. The different types of analyses presented emphasize once more the need for structured transparency information, which eventually can be used to improve transparency for all parties involved. 
To this end, we aim to provide more results about real-world data controllers, to investigate transparency information over time, or to compare the effects of different legislative frameworks, as soon as a significant corpus of documents exists.

Moreover, data protection authorities can play a pivotal role in promoting transparency by conducting audits and assessments of organizations' data protection practices. These assessments can serve as a basis for evaluating compliance with relevant regulations and identifying areas for improvement. The approach we presented above aims to support such efforts. To then further enhance transparency, data protection authorities can play an essential role by establishing comprehensive guidelines or frameworks for data controllers to follow. Such guidelines set clear standards for the format and content of transparency statements. While we demonstrate the general applicability of the Transparency Information Language and Toolkit (TILT), other similar structured representations can perform equivalently. We primarily argue for ensuring consistency and comparability within different industries and sectors. 

Our approach overcomes the limitations of previously presented related work, namely the P3P standard \cite{cranor2003p3p}. For instance, P3P only features a very limited vocabulary of purpose specifications, leading to imprecise policies. This lack of support limits its effectiveness in achieving its intended goals. Instead, the TAP allows for querying all the legally required information. To this end, our approach is consciously based on sharing practices being unveiled in comprehensive privacy policies. This also covers data sharing taking place in the backend, which cannot be observed from the client side. This is how our approach differs from cookie or tracker analysis and similar techniques explored in related work: While tools such as 
Disconnect\footnote{\url{https://disconnect.me/}},
Ghostery\footnote{\url{https://ghostery.com/}},
Thunderbeam/Lightbeam\footnote{\url{https://addons.mozilla.org/de/firefox/addon/lightbeam-chikl/}},
Cover Your Tracks (previously known as Panopticlick)\footnote{\url{https://coveryourtracks.eff.org/}},
or
Privacy Badger\footnote{\url{https://privacybadger.org/}}
 can be used to analyze the cookies set or requests made by a website, they do not provide information about the data-sharing practices happening behind the scence.
In contrast, our approach allows for a more comprehensive analysis of the data-sharing practices, including the sharing of personal data with third parties. Nevertheless, we consider our work a complement to the existing tools, as they can be used to analyze the data-sharing practices from the client side by laypersons, while our approach analyzes transparency information with more regulatory thoroughness. 

In addition to the network dynamics shown, more in-depth analyses could examine the exact relationships between different controllers. For instance, the GDPR distinguishes between joint controllerships (Art.~26), controller--processor relationships (Art.~27), %
different types of third-country transfers (Art.~44ff.), and others. Investigating these gets even more complex as soon as transfers happen not directly but through arbitrary sequences of legal entities. In the same vein, existing transparency statements remain vague about the transitive implications (e.g., regarding the purpose specification) (see also \cite{breitbarth2021risk}). Our platform, in turn, can be used as a starting point to model and further investigate these remaining questions. Similar research can be inspired by the ongoing analyses of privacy policies regarding their understandability or vagueness \cite{wagner2023privacy}. For such future work, we enable flexible interoperability through a dynamic API.

Further, the presented analysis methods not only foster transparency but also facilitate the enactment of other data subject rights. For example, for exercising the right to access (Art.~15), it is necessary to know which parties involved are processing personal data relating to a data subject, the responsible contact person, and information about how to retrieve a copy of said data (ex-post transparency, \cite{leschke2023streamlining}). Similarly, the TAP could support the right to rectification (Art.~16), erasure (Art.~17), or the restriction of processing (Art.~18), and others. The relevant information needed could already be (partially) queried from the existing graph structure. 

To expand on our findings, our future research aims to delve into a bigger number of real-world data controllers, investigate transparency information over time, and compare the effects of different legislative frameworks as soon as a significant corpus of transparency documents becomes available. In this line of research, we also want to investigate usability and efficacy of the transparency analysis platform in practice. Gathering feedback from more data protection professionals and stakeholders will help us refine our general approach, as well as the functionalities of the user or programming interfaces, ensuring they meets the needs of those involved in transparency assessments. Furthermore, such an evaluation has to comprise considerations of the underlying data quality, qualitative assessments of regulators' needs, and data subjects' expectations, adaptability to different legal frameworks, and the overall usability of the platform. The outcomes of such an evaluation should feed in the periodical review of regulatory frameworks, such as the expected major review of the GDPR in 2024.

Clearly, the quality of the analysis results is contingent upon the quality of the underlying transparency information (in our case TILT documents themselves) as input. Thus, it remains crucial to ascertain the extent to which the transparency information meets standards of completeness, accuracy, and timeliness, among other relevant factors. This limitation is also true for traditional privacy policies, which is why additional organizational measures, such as sample checks and quality assurance, should be defined and carried out by the supervisory authorities. Related work aims to establish a human-in-the-loop approach to create machine-readable transparency information with AI-generated text classification suggestions based on large language models. \cite{tiltify}

Lastly, data subjects should be provided with understandable summaries, visualizations, or other forms of effective communication about the relevant processing activities. Our prototype focused on the possibility to extract the necessary information, but could easily be extended to display the results in more versatile ways to better inform data subjects. In this regard, the presentation should be context-adaptive, preference-based, and competence-oriented \cite{gruenewaldEnablingVersatile}. Scholars identified the need for a more user-centric approach to transparency, which is why we aim to further investigate the needs of data subjects in future work upon the transparency analysis platform. To avoid the risk of information overload for services with a large number of data-sharing relationships, we also want to investigate the possibility of providing data subjects with a personalized view of the data-sharing network. This avoids or at least partly relaxes the \enquote{transparency paradox} \cite{nissenbaum2011contextual}, where more (transparency) information can lead to less understanding. However, supervisory authorities should still be able to access the full data-sharing network to conduct their assessments.

To sum up, we technically enable higher-level transparency analyses through a graph-structured representation and analysis of machine-readable transparency information in line with regulatory givens. Altogether, our contributions pave the way for future transparency research with points of reference not only for legal scholars and technical privacy engineers.

\section*{Acknowledgements}
{\small We thank Dennis Dittmann for some fruitful discussion on data-sharing networks and their analysis.}

\bibliographystyle{IEEEtran}
\bibliography{bibliography}

\end{document}